\documentclass{article}
\usepackage{spconf,amsmath,graphicx}

\usepackage{amsfonts}
\usepackage{booktabs}
\usepackage{caption}
\usepackage{subfig}

\usepackage{xcolor}

\title{Contrastive Unsupervised Learning for Speech Emotion Recognition}
\name{
	\begin{tabular}{c}
		Mao Li$^{\star}$, Bo Yang$^{\dagger}$, Joshua Levy$^{\dagger}$, Andreas Stolcke$^{\dagger}$, Viktor Rozgic$^{\dagger}$, Spyros Matsoukas$^{\dagger}$,\\Constantinos Papayiannis$^{\dagger}$, Daniel Bone$^{\dagger}$, Chao Wang$^{\dagger}$
	\end{tabular}
	}
\address{$^{\star}$Department of Computer Science, University of Illinois at Chicago \\
	$^{\dagger}$Amazon Alexa \\
$^{\star}$mli206@uic.edu, $^{\dagger}$\{amzbyang, levyjos, stolcke, rozgicv, matsouka, papayiac,\\danibone, wngcha\}@amazon.com}

\begin{document}
%
\maketitle
\begin{abstract}

Speech emotion recognition (SER) is a key technology to enable more natural human-machine communication. 
However, SER has long suffered from a lack of public large-scale labeled datasets. 
To circumvent this problem, we investigate how unsupervised representation learning on unlabeled datasets can benefit SER.
We show that the contrastive predictive coding (CPC) method can learn salient representations from unlabeled datasets, 
which improves emotion recognition performance.
In our experiments, this method achieved state-of-the-art concordance correlation coefficient (CCC) performance for all emotion primitives (activation, valence, and dominance) on IEMOCAP.
Additionally, on the MSP-Podcast dataset, our method obtained considerable performance improvements compared to baselines. 
\end{abstract}
\section{Introduction}
\label{intro}

Speech emotion recognition (SER) aims at discerning the emotional state of a speaker, 
thus enabling more human-like interactions between human and machines. 
An agent can understand the command of a human better if it is able to interpret the emotional state of the speaker as well. 
Moreover, a digital assistant can prove to be a human-like companion when equipped with the capability of recognizing emotions. 
These fascinating applications provide key motivations underpinning the fast growing research interest in this area \cite{trigeorgis2016adieu, schuller2018speech}
 
 Despite the substantial interest from both academia and industry, 
 SER has not found many real-world applications. 
 One possible reason is the unsatisfactory performance of existing systems. 
The difficulty is caused by, and contributes to, relatively small public data sets \cite{CarMurChi08_iemocap, RezCar19_msp} in this domain.
 The lack of large scale emotion annotated data hinders the application of deep learning methods, 
 from which many other speech-related tasks (e.g automatic speech recognition \cite{yu2016automatic}) have benefited greatly.

 In order to circumvent the data sparsity issue of SER, we investigate the use of unsupervised pre-training.
 Unsupervised pre-training techniques have received increased attention over the last few years. 
 The research interest in this direction is well-motivated: while deep-learning (DL) based methods achieve state-of-the-art 
 results across multiple domains, these methods tend to be data-intensive. 
 Training a large and deep neutral network usually requires very large labeled datasets. 
 The cost of data labeling has since become a major obstacle for applying DL techniques to real-world applications, and SER is no exception. 
Motivated by recent developments in unsupervised representation learning, we leverage an unsupervised pre-training approach for SER.

The proposed method shows great performance improvement on two widely used public benchmarks. 
The improvements on recognizing valence (positivity/negativity of the tone of voice) are particularly encouraging, as valence is known to be very hard to predict from speech data alone, see e.g. \cite{hanjalic2006extracting, mower2009interpreting}.
Furthermore, our analysis implies, even without explicit supervision in training, emotion clusters emerge in the embedding space of the pre-trained model, confirming the suitability of  unsupervised pre-training for SER.

\section{Related work}
\label{relate}

Recent studies on unsupervised representation learning have achieved great success in natural language processing \cite{TomBenMan20_gpt3,DevMinKen19_bert} 
and computer vision \cite{HeFanWu19_moco,AroYazOri18_cpc}.
While leveraging unsupervised learning for SER has been investigated relatively little, previous attempts using autoencoders have been successful \cite{eskimez2018unsupervised, deng2014introducing}.  More recently, it has been shown that learning to predict future information in a time series is a useful pre-training mechanism \cite{lian2019unsupervised}.  


Unsupervised methods based on contrastive learning have established strong and feasible baselines in many domains, recently.
For instance, contrastive predictive coding (CPC) \cite{AroYazOri18_cpc} is able to extract useful representations from sequential data
and achieves competitive performance on various tasks, including phone and speaker classification in speech. 
Our work relies on the use of a CPC network for learning acoustic representations from large unlabeled speech datasets.

\section{Background}
The primary goal of this study is to learn representations that encode emotional attributes shared across frames of speech audios without supervision.
We start by reviewing relevant concepts in emotion representation, then we give a brief review of the contrastive predictive coding (CPC) method.

 \subsection{Emotion representation}
In general, there are two widely used approaches to represent emotion: by emotion categories (happiness, sadness, anger, etc.) or by dimensional emotion metrics (aka emotion primitives) \cite{CarMurChi08_iemocap, RezCar19_msp, cowie2003describing}. Albeit intuitive, the categories-based representation may miss the subtleties of emotion ``strength'', e.g. annoyance versus rage. The dimensional emotion metrics often include activation (aka arousal, very calm versus very active) , valence (level of positivity or negativity) and dominance (very weak versus very strong).  In this work, we mainly focus on predicting dimensional emotion metrics from speech. Since emotion representation is an active research topic, we refer the interested readers to \cite{cowie2003describing, yannakakis2018ordinal}.


\subsection{Contrastive predictive coding }



As the name suggests, CPC falls into the contrastive learning paradigm: positive example and negative examples are constructed, and the loss function encourages separation of positive from negative examples. We give a detailed description of CPC below.

For an audio sequence $X = (x_1, x_2, ..., x_n)$, 
CPC uses a nonlinear encoder $f$ to project observation $x_t \in \mathbb{R}^{D_x}$ to its latent representation $z_t = f(x_t),\text{ where } z_t \in \mathbb{R}^{D_z}$.
Then an autoregressive model $g$ is adopted to aggregate $t$ consecutive latent representations from the past into a contextual representation $c_t = g(z_{\le t}), \text{ where } c_t \in \mathbb{R}^{D_c}$.

Since $c_t$ summarizes the past, it should be able to infer the latent representation $z_{t+k}$ of future observations $x_{t+k}$ from $c_t$, for a small $k$. For this purpose, 
a prediction function $h_k$ for a specific $k$ takes the context representation as the input to predict the future representation:
\begin{align}
    \hat{z}_{t+k} = h_k(c_t) = h_k(g(z_{\le t})).
\end{align}
To form a contrastive learning problem, some negative samples (i.e. other observation $x$) are drawn, either from the same sequence or other sequences, and their latent representations ($z$) are computed.

Assuming $N-1$ negatives are randomly sampled for each context representation,
then positive and negatives form a set of $N$ samples that contains only one positive and $N-1$ negatives.
To guide feature learning, the CPC method proposes to discriminate the positive from negatives, which boils down to an N-way classification problem.
CPC uses the infoNCE loss function: for an audio segment and a time step $t$, the infoNCE loss is defined as
\begin{align}
    \mathcal{L}= - \sum_{m=1}^k \left[ \log \frac{\exp(\hat{z}_{t+m}^\top z_{t+m})/\tau}{\exp(\hat{z}_{t+m}^\top z_{t+m})/\tau + \sum_{i=1}^{N-1}\exp(\hat{z}_{t+m}^\top z_{i})/\tau} \right],
    \label{eq:nceloss}
\end{align}
where $\tau$ is a scaling factor (a.k.a temperature) to control the concentration-level of the feature distribution,
$k$ is the upperbound on time extrapolation. Notice that the summation over $i$ assumes that the randomly drawn negative samples are labeled as $\{1,...N-1\}$, and these are different for each $z_{t+m}$. In addition, the loss function considers all the future time extrapolation up to $k$. Clearly, the loss \eqref{eq:nceloss} is additive across different audio segments and time steps, hence in training, the loss \eqref{eq:nceloss} is usually computed for batches of audio segments and all possible time steps in these segments, to utilize the mini-batch-based Adam \cite{kingma2014adam} optimizer.

Optimizing (\ref{eq:nceloss}) results in larger inner product between a latent representation and its predicted counterpart, than any of the negatives --  mismatched latent representation and predictions. Theoretical justification for the optimization objective function \eqref{eq:nceloss} can be found in \cite{AroYazOri18_cpc} and \cite{BenSheAra19_mi}.

\section{Proposed method}
\label{method}
The proposed method consists of two stages: pre-training a ``feature extractor'' model with CPC on a large un-labeled dataset, and training an emotion recognizer with features learned in the first stage. In this section, we introduce the emotion recognizer and training loss function.
\subsection{Attention-based emotion recognizer}
\label{attention-recognizer}

The output of CPC is a sequence of encoded vectors $C = \{c_1, c_2, ..., c_L\}, C\in \mathbb{R}^{L\times D_c}$.
To predict primitive emotions for a certain speech utterance, an utterance-level embedding is desired.
Since certain parts of an utterance are often more emotionally salient than others,
we adopt a self-attention mechanism to focus on these periods for utilizing relevant features.
Specifically, a structured self-attention \cite{vaswani2017attention} layer aggregates information from the output of CPC 
and produces a fixed-length vector $u$ as the representation of the speech utterance.

Given $C$ as input of the emotion recognizer, we follow \cite{vaswani2017attention} to compute the scaled dot-product attention representation $H$ as
\begin{align}
    H &= \text{ softmax }\left(CW_Q (CW_K)^\top / \sqrt{D_{attn}}\right)CW_V
\end{align}
where $W_Q$, $W_K$, and $W_V$  are trainable parameters, and all have shape $D_c \times D_{attn}$. The subscripts $Q, K, V$ stand for query, key, and value, as defined in \cite{vaswani2017attention}.

In order to learn an embedding from multiple aspects, we use a multi-headed mechanism to process the input multiple times in parallel.
The independent attention outputs are simply concatenated and linearly transformed to get the final embedding $U\in \mathbb{R}^{D_u}$.
\begin{align}
    H^j &= \text{ softmax }\left(W_Q^jC(W_K^jC)^\top / \sqrt{D_{attn}}\right)W_V^jC \\
    U   &= \text{Concat}(H^1, H^2, ..., H^n)W_{O} 
\end{align} 
where $W_{O}\in\mathbb{R}^{nD_{attn}\times D_u}$ is another trainable weight matrix, and $U \in \mathbb{R}^{L\times D_u}$ is the sequence representation after the multi-headed attention layer.

Following the multi-headed attention layer, we compute the mean and standard deviation along the time dimension, and concatenate them as the sequence representation
\begin{align}
	u = [\text{ mean }(U); \text{ std }(U)]
\end{align}
Subsequently, two dense layers with ReLU activation are used.
We apply a dropout after these two dense layers with a small dropout probability.
The final output layer is a dense layer with hidden units of the number of emotion attributes 
(e.g. three dimensions corresponding to activation, valence and dominance respectively).

\subsection{Loss function}
Following \cite{weninger2016discriminatively}, we build a loss function based on the concordance correlation coefficient (CCC, \cite{lawrence1989concordance}). For two random variables $X$ and $Y$, the CCC is defined as
\begin{align}
	\text{CCC}(X, Y) = \rho\frac{2\sigma_X\sigma_Y}{\sigma_X^2 + \sigma_Y^2 + (\mu_X - \mu_Y)^2},
	\label{eq:ccc}
\end{align}
where $\rho = \frac{\sigma_{XY}}{\sigma_X\sigma_Y}$ is the Pearson correlation coefficient, and $\mu$ and $\sigma$ are the mean and standard deviation, respectively. As can be seen from \eqref{eq:ccc}, CCC measures alignment of two random variables. In our setting, model predictions and data labels assume the role of $X$ and $Y$ in \eqref{eq:ccc}.

Since the emotion recognizer predicts at the same time activation, valence and dominance, we use a loss function that combines $\text{CCC}_{act}$, $\text{CCC}_{val}$, $\text{CCC}_{dom}$ values for activation, valence, and dominance, respectively
\begin{align}
    \mathcal{L} = 1 - \alpha \text{CCC}_{act} - \beta \text{CCC}_{val} - \gamma\text{CCC}_{dom}
\end{align}
We set the trade-off parameters $\alpha = \beta = \gamma = 1/3$ in all our experiments. 

\section{Speech corpora }
\label{data}

For unsupervised pre-training, we train the CPC model on LibriSpeech dataset \cite{VasGuoDan15_Librispeech},
which is a large scale corpus originally created for automatic speech recognition (ASR).
It contains 1000 hours of  English audiobook reading speech, sampled at 16kHz.
In our experiment, due to computational limitations, we use an official subset "train-clean-100" containing 100 hours of clean speech for unsupervised pre-training. 
In this subset, 126 male and 125 female speaker were assigned to the training set. For each speaker, the amount of speech was limited to 25 minutes to avoid imbalances in per-speaker duration.

To evaluate the empirical emotion recognition performance, 
we perform experiments on the widely used MSP-Podcast dataset \cite{RezCar19_msp} and IEMOCAP dataset \cite{CarMurChi08_iemocap}.
MSP-Podcast is a database of spontaneous emotional speech.
In our work, we used version 1.6 of the corpus, which contains 50,362 utterances amounting to 84 hours of audio recordings.
Each utterance contains a single speaker with duration between 2.75s and 11s. We follow the official partition of the dataset, which has 34,280, 5,958, and 10,124 utterances in the training, validation and test sets, respectively. The dataset provides scores for activation, valence and dominance, as well as categorical emotion labels.

IEMOCAP is a widely used corpus in SER research. It has audio-visual recordings from five male and five female actors.
The actors were instructed to either improvise or act out certain specific emotions. The dataset contains 5,531 utterances grouped into 5 sessions, which amount to about 12 hours of audio.
Similar to MSP-Podcast, this dataset  provides categorical and dimensional emotion labels. In this work, we focus on predicting the dimensional emotion metrics from the speech data.





\section{Experiment results}
\label{exp}

\subsection{Setups}
Our experiments investigate four different setups:

\noindent
\textbf{a). supervised only (Sup)}: As a simple baseline, an emotion recognizer was trained and tested on 40-dimensional log filterbank energies (LFBE) features of IEMOCAP and MSP-Podcast, respectively.
LFBE features have been tested in a wide variety of applications.

\noindent
\textbf{b). joint CPC + supervised (jointCPC)}: JointCPC trained CPC model and emotion recognizer in an end-to-end manner, where the CPC model aims to learn features from the raw audios directly, while the Sup setup uses hand-crafted features for the supervised task.
    We included this baseline to test whether it is possible to learn better features  when the feature extraction part is aware of the downstream task.

\noindent
\textbf{c). miniCPC}: Compared with jointCPC, miniCPC trains the CPC model and the emotion recognizer in two separate stages on the same datasets. 
    In this setup, we can verify whether CPC model can learn universal representations that can facilitate various downstream tasks. 
    
\noindent
\textbf{d). CPC pre-train + supervised (preCPC)}: We first pretrained a CPC model with a 100-hour subset of the LibriSpeech dataset. 
    Then an attention-based emotion recognizer will be trained on features that were extracted from the learned CPC model with MSP-Podcast and IEMOCAP, respectively. 
    Since the training corpus for CPC is much larger than the labeled datasets, we can test whether introducing a large out-of-domain dataset for unsupervised pretraining is useful.

For the CPC model used in the above settings,
we use a four layer CNN with strides [5, 4, 4, 2], filter-sizes [10, 8, 8, 4] 
and 128 hidden units with ReLU activations to encode the 16KHz audio waveform inputs.
A unidirectional gated recurrent unit (GRU) network with 256 hidden dimensions is used as the autoregressive model.
For each output of GRU, we predict 12 timesteps in the future using 50 negative samples, sampled from the same sequence, in each prediction.
We train the CPC model with fixed length utterances of 10s duration.
Longer utterances are cut at 10s, and shorter ones were padded by repeating themselves.

For the emotion recognizer, an 8-head attention layer with 512 dimensional hidden states is used.
The outputs of attention layer have the same dimension of the inputs.
The two fully-connected layers have 128 hidden units.
The drop out probability is set to 0.2 for the dropout layers.

Our model was implemented in PyTorch and all methods were conducted on 8 GPUs each with a minibatch size of 8 examples for CPC pretraining. 
We use Adam optimizer with a weight decay of 0.00001 and a learning rate of 0.0002.
We used 50 epochs for training and saved the model that perform best on validation set for testing.

To evaluate the IEMOCAP dataset, we configured 5-fold cross-validation to evaluate the model.
All experiments were run five times to produce the means and standard deviations.

\subsection{Results}
Table \ref{tab:iemocap} and \ref{tab:msp} present the performance in terms of CCC for activation, valence and dominance on the IEMOCAP and MSP-Podcast corpora, respectively.
As shown in these tables, on both datasets preCPC consistently outperforms other setups.
preCPC achieves higher CCC values for all metrics than Sup, which implies that the representations learned by CPC are superior to hand-crafted features for speech emotion recognition task.  { Surprisingly, even pre-training the CPC model on a small dataset, miniCPC still performs better than jointCPC on both datasets.
We hypothesize that this is because unsupervised pre-training learns universal representations that are less specialized towards solving a certain task.
Hence, it produces representations with better generalization which might facilitate various downstream tasks.
However, for the jointCPC method, a trade-off has to be made between emotion prediction capability and representation learning.}
Also notice that, preCPC outperforms miniCPC by a large margin. 
This confirms our intuition that exposing the model to more diverse acoustic conditions and speaker variations is beneficial for learning robust features.

\begin{table}[t!]
    \caption{CCC scores (mean/std) on the IEMOCAP dataset}
    \centering
    \label{tab:iemocap}
    
    \begin{tabular}{@{}rcccc@{}}
        \toprule
        Methods & $\text{ CCC }_{\text{avg}}$ & $\text{ CCC }_{\text{act}}$ & $\text{ CCC }_{\text{val}}$ & $\text{ CCC }_{\text{dom}}$ \\
        \midrule
        Sup      & .664 {\scriptsize $\pm$ .007} & .638 {\scriptsize $\pm$ .017} 
                & .718 {\scriptsize $\pm$ .004} & .635 {\scriptsize $\pm$ .009} \\
        jointCPC  & .562 {\scriptsize $\pm$ .012} & .549 {\scriptsize $\pm$ .032} 
                & .642 {\scriptsize $\pm$ .013} & .491 {\scriptsize $\pm$ .016} \\
        miniCPC & .660 {\scriptsize $\pm$ .005} & .673 {\scriptsize $\pm$ .028} 
                & .702 {\scriptsize $\pm$ .009} & .606 {\scriptsize $\pm$ .019} \\
        preCPC     & .731 {\scriptsize $\pm$ .003} & .752 {\scriptsize $\pm$ .014} 
                & .752 {\scriptsize $\pm$ .009} & .691 {\scriptsize $\pm$ .009} \\
        \bottomrule
    \end{tabular}

\end{table}

\begin{table}[t!]
    \caption{CCC scores (mean/std) on the MSP-Podcast dataset}
    \centering
    \label{tab:msp}
    \begin{tabular}{@{}rcccc@{}}
        \toprule
        Methods & $\text{ CCC }_{\text{avg}}$ & $\text{ CCC }_{\text{act}}$ & $\text{ CCC }_{\text{val}}$ & $\text{ CCC }_{\text{dom}}$ \\
        \midrule
        Sup      & .458 {\scriptsize $\pm$ .005} & .596 {\scriptsize $\pm$ .007} 
                & .266 {\scriptsize $\pm$ .004} & .501 {\scriptsize $\pm$ .013} \\
        jointCPC  & .491 {\scriptsize $\pm$ .008} & .628 {\scriptsize $\pm$ .006} 
                & .280 {\scriptsize $\pm$ .006} & .568 {\scriptsize $\pm$ .007} \\
        miniCPC & .549 {\scriptsize $\pm$ .006} & .688 {\scriptsize $\pm$ .009} 
                & .345 {\scriptsize $\pm$ .005} & .615 {\scriptsize $\pm$ .011} \\
        preCPC     & .571 {\scriptsize $\pm$ .004} & .706 {\scriptsize $\pm$ .006} 
                & .377 {\scriptsize $\pm$ .008} & .639 {\scriptsize $\pm$ .012} \\
        \bottomrule
    \end{tabular}

\end{table}

    
\begin{figure}[t!]
        \centering
        \includegraphics[width=0.7\linewidth]{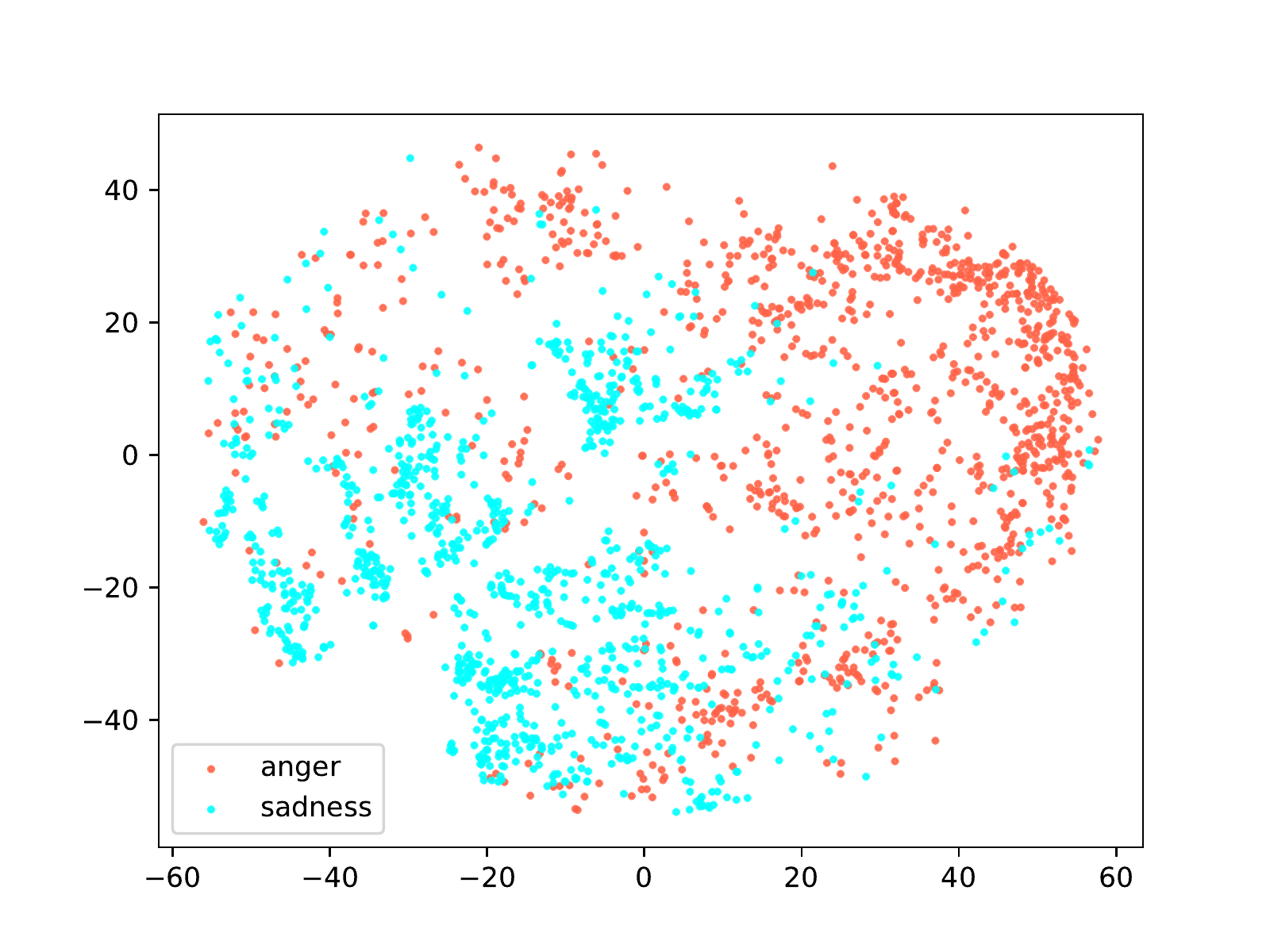}
        \caption{Visualization of the learned representations}
        \label{fig:tsne_2class}
\end{figure}

We also plot the representations extracted by CPC from IEMOCAP to examine how suitable these representations are for emotion. For visualization purposes, we used the categorical emotion labels when making the figure.
As can be seen from Figure  \ref{fig:tsne_2class},  the CPC model representation is capable of separating sadness from anger to a good extent, even though it is trained without emotion labels.

\section{Conclusion}
\label{conclusion}

Our experiment results demonstrated that CPC can learn useful features from unlabeled speech corpora that benefit emotion recognition.
We have also observed 
significant performance improvement on widely used public benchmarks under various experiments setups,
compared to baseline methods.
Further, we also present a visualization that confirms the discriminative nature, with respect to emotion classes, of the CPC-learned representations.

So far we mainly conducted experiments on LibriSpeech for pre-training.
In the future, it would be interesting to investigate the impact of other corpora for pre-training.
In particular, corpora that have more varied and expressive emotions might yield representations that are even more relevant for SER. 

\vfill\pagebreak


\bibliographystyle{IEEEbib}
\bibliography{strings,refs}

\end{document}